\documentclass[conference]{IEEEtran}
\usepackage{cite}
\usepackage{amsmath,amssymb,amsfonts}
\usepackage{algorithmic}
\usepackage{graphicx,color}
\usepackage{textcomp}
\usepackage{multirow}
\usepackage{enumitem}
\usepackage{caption}
\usepackage{subcaption}
\usepackage{lipsum}
\usepackage{colortbl}
\usepackage{diagbox}

\def\BibTeX{{\rm B\kern-.05em{\sc i\kern-.025em b}\kern-.08em
    T\kern-.1667em\lower.7ex\hbox{E}\kern-.125emX}}
\begin{document}

\title{Exploring High Tower Base Stations with Multi-User Massive MIMO for Rural Connectivity}



\author{\IEEEauthorblockN{Ammar El Falou, Mohamed-Slim~Alouini}\\
 \IEEEauthorblockA{Computer, Electrical and Mathematical Science and Engineering (CEMSE) Division \\ King Abdullah University of Science and Technology (KAUST) \\ 23955 Thuwal, Saudi Arabia}
Email: \{ammar.falou, slim.alouini\}@kaust.edu.sa}

\begin{titlepage}
\vspace{3cm}
\Huge\noindent\textbf{IEEE Copyright Notice}\\
\vspace{3cm}

\large\noindent\copyright 2023 IEEE. Personal use of this material is permitted. Permission from IEEE must be obtained for all other uses, in any current or future media, including reprinting/republishing this material for advertising or promotional purposes, creating new collective works, for resale or redistribution to servers or lists, or reuse of any copyrighted component of this work in other works.
\end{titlepage}

\maketitle

\begin{abstract}
The digital divide is a key issue worldwide. Almost $3$ billion people, mainly in rural areas, are still not connected. In this paper, we explore the capability of high towers base station (HTBS) with massive multiple input multiple output (mMIMO) in offering low-cost rural connectivity. We previously showed the benefits of HTBS in the downlink. We focus in this work on the uplink (UL) where we compute the UL data rate per user for different values of transmit effective isotropic radiated power (EIRP). Our results show that the HTBS solution is viable as relatively good user UL rates are achieved with reasonable EIRPs. This is of high interest for covering rural areas, characterized by low population densities and a low number of active users, as the coverage is their main constraint, rather than the capacity as in urban areas. Techno-economical aspects such as the recommended frequency for HTBS, the number of covered persons, the average population density of covered rural areas, and potential low-cost locations for HTBS are provided. Non-technological challenges for the HTBS solution are also discussed. 
\end{abstract}



\IEEEpeerreviewmaketitle

\vspace{-0.1cm}
\section{Overview}
The digital divide is a prevailing issue worldwide. On the one hand, more than half of humanity is benefiting from the services provided by modern communication systems in their daily life. On the other hand, almost 3 billion people are still not connected~\cite{yaacoub2020key,feltrin2021potential,ahmmed2022digital,falou2022enhancement}. These people generally reside in rural areas or undeveloped regions with bad infrastructure as a lack of roads, regular power outages, and a scarcity of skilled engineers. The low number of potential customers in poor countries and/or rural areas has made legacy connectivity solutions non-viable. Indeed, the cost of cellular network deployment and its operational expenditures (OPEX) exceeds potential revenues based on affordable user subscription fees.      

Under the recent efforts in 6G to decrease the digital divide, several solutions to provide ubiquitous connectivity are proposed~\cite{yaacoub2020key,ahmmed2022digital,zhang2021challenges,Castellanos2021,dang2021big,osoro2021techno,dang2020should,harri2021,feltrin2021potential,qin2022drone,falou2022enhancement}. They highlight that wireless is the only option for access and backhauling links. Contrary to wired communications, the deployment of wireless technologies does not require high capital expenditure (CAPEX) making them suitable for the limited expected revenues in rural areas. One can refer to satellite communications, high altitude platform system (HAPS), unmanned aerial vehicle (UAV), and high tower base station (HTBS). Among them, the HTBS solution is the most cost-effective, especially when deployed at the top of available towers such as TV towers, wind turbines, or high places such as hills, mountains, etc. \cite{yaacoub2020key,bondalapati2020supercell,taheri2021potential,feltrin2021potential,falou2022enhancement,matracia2021exploiting}.    

Thanks to their high beamforming gains, massive MIMO (mMIMO) techniques are expected to be a crucial part of 6G standards \cite{MassiveMIMOBook,dang2020should}. For mMIMO systems, multi-user (MU) precoding/combining enhance spatial selectivity and reduce intra-user interference \cite{khaled2021multi,khaled2023angle}. In \cite{falou2022enhancement}, we have explored the potentiality of HTBS with MU mMIMO to offer rural connectivity in the downlink (DL). We have shown that one HTBS can cover an area much larger than the one covered by a legacy BS. The target DL user rate was chosen to be at least $10$~Mbps considered as a sufficient value to guarantee a good quality of service for most of the applications~\cite{yaacoub2020key}. 
 
The extension of the coverage range in the DL will certainly impact the uplink (UL) user data rate. The increase in the distance between the user equipment (UE) and the BS increases the signal path loss. To complement our study in \cite{falou2022enhancement}, we evaluate in this work the UL data rate per user for different effective isotropic radiated power (EIRP) values. We show that for reasonable EIRPs, the obtained UL rates are acceptable and that the high DL coverage distances proposed in \cite{falou2022enhancement} can be maintained. We highlight that HTBSs can have many practical use cases: 
\begin{enumerate}
    \item Cover a wide non-covered rural area,
    \item Form a coverage umbrella of an existing cellular network. In most rural areas, if cellular networks exist, they are only deployed in villages and main roads. HTBS and legacy networks can use different frequencies, where the low-frequency band at HTBSs ensures coverage, and the high-frequency band at legacy BSs ensures a high rate for its respective users \cite{3GPPNR}, 
    \item In a more advanced UE, carrier aggregation between a legacy cellular network and HTBSs can be done\cite{3GPPNR}. This will permit the combination of the rates from HTBSs and legacy BSs.  
\end{enumerate}

Our work is important to motivate the deployment of HTBS in the effort of reducing the digital divide. The techno-economical aspects and non-technological challenges of the HTBS solution are also addressed.  

The remainder of this paper is organized as follows. Section~\ref{sec:SystemModel} presents the system model and the performance metric. Section~\ref{sec:NumericalResults} provides the numerical results. Section~\ref{sec:techno} discusses the techno-economical aspects. Section~\ref{sec:nontechno} provides the non-technological challenges. Conclusions are drawn in Section~\ref{sec:Conclusion}.


\section{System Description}
\label{sec:SystemModel}
\begin{figure}
    \centering
    \includegraphics[width=0.96\columnwidth]{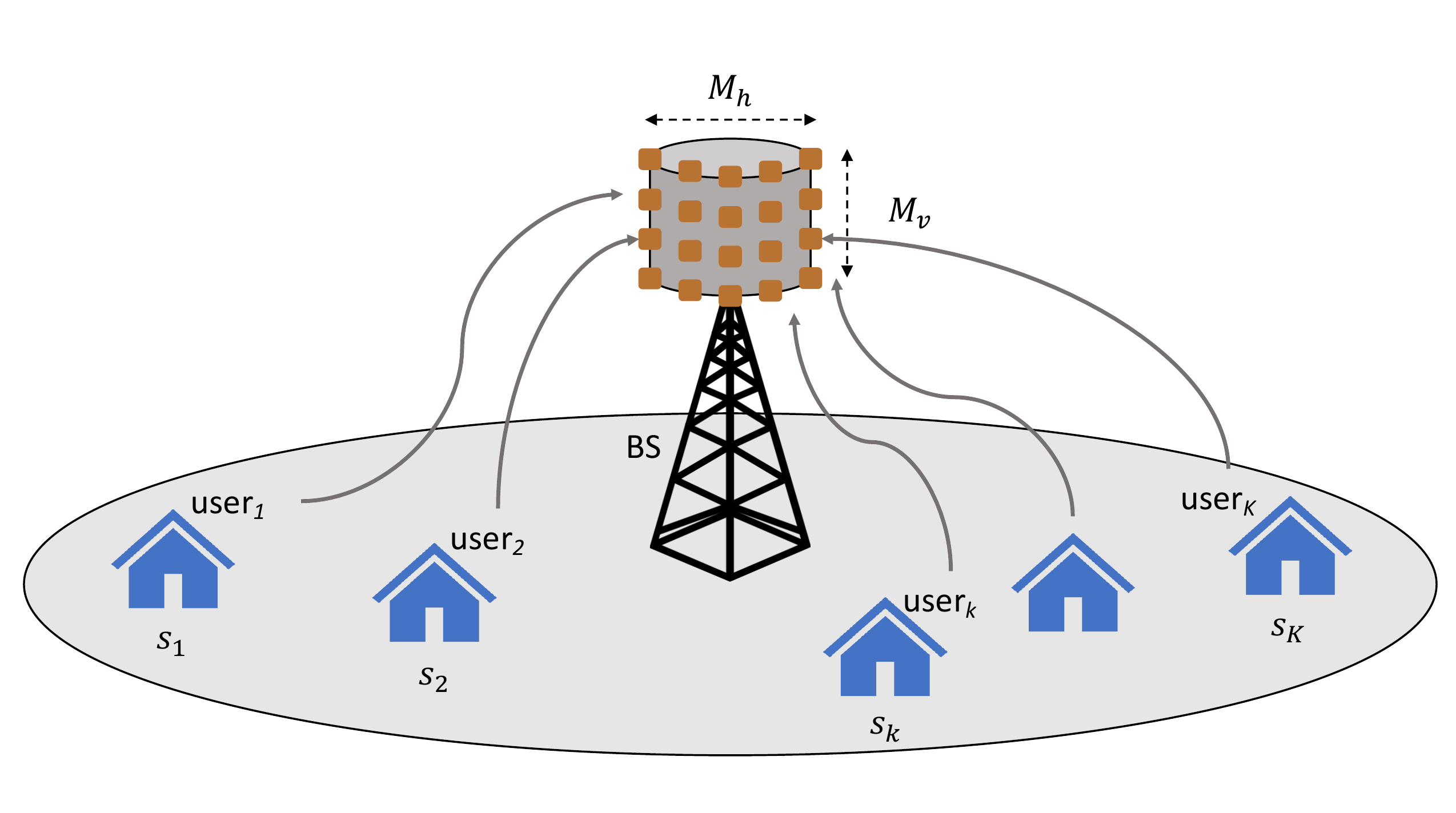}
    \caption{A massive MIMO base station with UCyA.}
    \label{fig:SystemModel}
\end{figure}






Similar to our work in \cite{falou2022enhancement}, we consider a MU mMIMO BS with $M = 512$ dual-polarized antennas ($256$ physical antennas) forming a uniform cylindrical array (UCyA), depicted in Fig.~\ref{fig:SystemModel}. This BS serves simultaneously $K$ single-antenna users. The ratio $M/K \geq 1$. In Fig.~\ref{fig:SystemModel}, $M_h$ denotes the number of horizontal antennas and $M_v$ 
denotes the vertical ones. 

\subsection{Uplink System Model}
At the BS, the received signal is equal to the summation of transmitted signals from the $K$ users affected by the channel coefficients and the noise
\begin{equation}
{\bf y} =  \underbrace{\sum_{k=1}^{K} {\bf h}_{k} { s}_{k}}_{\text{Desired signals }} 
+ \underbrace{{n}_{k}}_{\text{Noise}} ,
\label{eq:yk_UL}
\end{equation}
where $n_{k}$ $\sim N_\mathbb{C} ({\bf 0}_{M},\sigma^{2}{\bf I}_M)$ is the independent additive complex noise with zero mean and variance $\sigma^{2}$. ${\bf h}_{k}$ denotes the channel state information (CSI) between user $k$ and the BS. The UL signal from user $k$ is denoted by ${s}_{k} \in \mathbb{C}$ and has a power $p_k=\mathbb{E} \{|s_k|^2\}$ (we use $p_k$ and $\text{EIRP}_k$ interchangeably in this paper representing the effective transmit power at user $k$). The BS selects the receive combining vector ${\bf v}_{k} \in \mathbb{C}^{M}$ to separate its $k$-th desired user signal from the intra-cell interference as

\begin{equation}
{\bf v}_{k}^{H}{\bf y} = \underbrace{{\bf v}_{k}^H {\bf h}_{k} s_{k}}_{\text{Desired signal}} + 
\underbrace{\sum_{i=1, i \neq k}^{K} {\bf v}_{k}^H {\bf h}_{i} s_{i}}_{\text{Intra-cell interference}} 
+ \underbrace{{\bf v}_{k}^{H}{n}_{k}}_{\text{Noise}}.
\label{eq:vk_UL}
\end{equation}

\subsection{Channel Models}
In order to have realistic results, we consider the 3GPP channel models developed for rural environments, i.e., 3GPP rural macro (RMa) $38.901$ \cite{3GPP}. We differentiate between legacy BS and HTBS by selecting the non-line of sight (NLoS) model for the former, and the LoS model for the latter. The high height of the HTBS increases the probability of having a LoS between the HTBS and the user. The channel coefficients are generated using the geometry-based quasi-deterministic radio channel simulator (QUADRIGA) version 2.6~\cite{quad}. The users are randomly distributed around the BS. The BS is set in the cell center. The UCyA is installed at the top of the BS mast. The separation distance between antennas in the UCyA is equal to the half of wavelength, i.e., $\lambda/2$. Note that QUADRIGA accounts for several geometric aspects as the correlation between antennas, the location, and the height of the BS and users. 

\subsection{RZF Linear Receive Combining Scheme}
\label{sec:MIMOPrecoders}

\begin{table*}
\normalsize \centering 
\caption{Legacy BS and HTBS Parameters.} 
\begin{tabular}{|c|c|c|} 
\hline
Parameter & Legacy BS & HTBS \\
\hline
\multicolumn{3}{|c|}{} \\[-13pt]
\hline
User Height [m]    & $8$ & $8$ \\  
\hline
BS Height [m]    & $25$ & $150$ \\  
\hline
3GPP Channel Model     & 38.901 RMa NLOS & 38.901 RMa LOS \\       
\hline 
Nb. of Active Users $K$    & $20,50,100$ & $ 20,50,100$ \\
\hline
Nb. of Antennas $M$ (dual)   & $512$  & $512$  \\
\hline
 $M_h \times M_v$ (dual) & $32 \times 8\times2$ & $32 \times 8\times2$ \\
\hline
Bandwidth $W$ [MHz] &  $10, 20, 100$ & $10, 20, 100$ \\ 
\hline
Cyclic Prefix Overhead  & $5\%$ & $5\%$ \\
\hline
$\text{EIRP}_k$ [dBm] & $40, 33, 30, 23$ & $40, 33, 30, 23$ \\
\hline
Power Ratio $\delta$ [dB] & $20$ & $20$ \\
\hline
Carrier Frequency $f_c$ [MHz] & \begin{tabular}{lcr} $700$ & $1800$ & $3500$  \end{tabular} & \begin{tabular}{lcr}$700$&$1800$&$3500$ \end{tabular} \\

\hline 
\end{tabular}
\label{tab:SystemParmeters}
\end{table*}





In our work, the regularized zero-forcing (RZF) linear receive combining scheme is employed. The RZF is an efficient and low-complex combining scheme for MU mMIMO systems. The remaining linear combining schemes such as conjugate beamforming (CB), Zero Forcing (ZF), or minimum mean-squared error (MMSE) are not recommended. ZF combining is not suitable when users exhibit low signal-to-noise ratios (SNR). CB combining is low-complex but offers a limited data rate and is not resistant to channel estimation errors. Finally, MMSE combining is optimal but complex for implementation in practice. In our work, we assume a perfect CSI. Fortunately, the RZF combining scheme provides good performance even when using low complex estimators as least-squares (LS)~\cite{MassiveMIMOBook}. 

In order to give the expression of the RZF receive combining scheme,  we define the $M \times K$ matrix ${\bf H}$ having all channel coefficients between the $K$ users and the antennas of the BS 
    \begin{equation}
    {\bf H}=[{\bf h}_{1} \dots {\bf h}_{K}] .
    \end{equation}
    


    
  

The RZF receive combining scheme matrix is given by
\begin{equation}
       {\bf V}^\text{RZF} = {\bf H}\left(({\bf H})^H {\bf H}  + \sigma^{2} {P}^{-1} \right)^{-1} ,
    \end{equation}
where ${P} = \text{diag}(p_1,...,p_K)$ is the diagonal matrix of the $p_k$ of all active users in the cell and the receive combining vector for each user ${\bf v}_{k} \in \mathbb{C}^{M}$ is the column of $ {\bf V}^\text{RZF}$ as
\begin{equation}
{\bf V}^\text{RZF} = \left[{\bf v}_{1} \dots {\bf v}_{K} \right].
\end{equation}  

To have a realistic analog-to-digital converter at the BS, the received power ratio between the weakest and the strongest users, denoted by $\delta$, is to be limited \cite{MassiveMIMOBook}. This will imply a limitation of the $\text{EIRP}_k$ of each user. 

\subsection{Performance Metric}


In the UL, the spectral efficiency $\text{SE}_{k}$ [bit/s/Hz] of user $k$ is given by \cite{MassiveMIMOBook}
\begin{equation}
\label{eq:SE_UL}
    \text{SE}_{k} = \frac{\tau_u}{\tau_c} \mathop{\mathbb{E}} \{\log_2(1+\text{SINR}_{k}) \}  ,
\end{equation}
where $\text{SINR}_{k}$ denotes the signal-to-interference-plus-noise ratio of user $k$ and is given by
\begin{equation}
\label{eq:SINR_UL}
  \text{SINR}_{k}= \frac{p_{k} |{\bf v}^{H}_{k} {\bf h}_{k}|^2}{\sum_{i=1, i \neq k}^{K} {p_{i} |{\bf v}^{H}_{k} {\bf h}_{i}|^2} + \sigma^2},
\end{equation}
$\frac{\tau_u}{\tau_c}$ represents the portion of samples used for uplink data transmission per coherence block $\tau_c$. For frequency division duplex (FDD) mode, the total bandwidth is used for the UL, while for time division duplex (TDD) $\frac{\tau_u}{\tau_c}$ proportion of the bandwidth is used for UL. The data rate per user $\text{R}_{k}$ is obtained by the multiplication of the $\text{SE}_{k}$ with the system bandwidth $W$ as
\begin{equation}
\label{eq:data_rate}
\text{R}_{k} =W\times\text{SE}_{k} .
\end{equation}
 
In a real system, $\tau_p$ samples are used for channel estimation. For a TDD system and assuming channel reciprocity in the UL and DL, $\tau_p = K$. For a FDD system, $\tau_p = M$ in the DL and $\tau_p = K$ in the UL. In order to obtain realistic results, we considered the effect of $\tau_p$ on $\text{SE}_{k}$, where the estimation samples are subtracted from $\tau_c$.

\subsection{Types of Base Stations}
We differentiate between two types of BSs:
\begin{itemize}
    \item HTBS: with an antenna array at $150$~m height, typical for TV towers in rural lands~\cite{ITU}.   
    \item Legacy BS: with an antenna array at $25$~m, considered as a benchmark. 
\end{itemize}

As mentioned earlier, HTBS can have many practical use cases. In the co-existence cases of HTBS and legacy BS, we assume that they are not using the same frequency and that there is no interference between legacy BS and HTBS. 

\begin{table*}
\normalsize \centering 
\caption{Achievable UL rate per user $\text{R}_{k}$ in [Mbps] for a Legacy BS and a HTBS.} 
\setlength\tabcolsep{3pt}
\begin{tabular}{|c|c|c|c|c|c|c|c|} 
\hline
\rule{0pt}{10pt}    
 Type & $K$ & $f_c$ [MHz] & $d_\text{cov}$ [km] & $\text{EIRP}_k =40$ [dBm] & $\text{EIRP}_k=33$ [dBm]  & $\text{EIRP}_k=30$ [dBm] & $\text{EIRP}_k=23$ [dBm] \\
\hline
\multicolumn{8}{|c|}{} \\[-13pt]
\hline
\multirow{9}{*}{Legacy BS} & $20 $&  & $4.9 $& $27 $& $12 $& $8 $&  $2.3 $\\  
\cline{2-2}\cline{4-8} 
 & $50 $& $700 $& $3.4 $& $30$& $14.5 $& $10$& $3.2 $\\  
\cline{2-2}\cline{4-8}
 & $100 $& (FDD) & $2.1 $& $39 $& $21 $& $15 $&  $5.7 $\\  
\cline{2-8}
\multicolumn{8}{|c|}{} \\[-11.5pt]
\cline{2-8}
 & $20 $&  & $3.5 $& $28 $& $9.7 $& $5.6 $&  $1.4 $\\  
\cline{2-2}\cline{4-8}
 & $50 $& $1800 $& $2.5$& $34 $& $13.3 $& $8 $& $2.4 $\\  
\cline{2-2}\cline{4-8}
 & $100 $& (FDD) & $1.6$& $41 $& $16.2 $& $10.7 $&  $4.3 $\\  
\cline{2-8}
\multicolumn{8}{|c|}{} \\[-11.5pt]
\cline{2-8}
 & $20 $&  & $2.7 $& $8.86$& $2.3 $& $1.3 $&  $0.3$\\  
\cline{2-2}\cline{4-8}
 & $50 $& $3500 $&$1.9 $& $12 $& $3.7 $& $2.1 $& $0.64 $\\  
\cline{2-2}\cline{4-8}
& $100 $& (TDD) & $1.4$& $13.2 $& $4.5 $& $2.7 $& $0.93$\\  
\hline
\multicolumn{8}{|c|}{} \\[-13pt]
\hline
\multirow{9}{*}{\begin{tabular}{c} HTBS \\  \end{tabular}} & $20 $&  &$37 $& $12.5 $& $4.5 $& $2.5 $&  $0.85 $\\ 
\cline{2-2}\cline{4-8}
 & $50 $& $700 $&$21 $& $16.5 $& $7 $& $4.5 $& $1.55 $\\  
\cline{2-2}\cline{4-8}
 & $100 $& (FDD) & $12.5$& $22 $& $10 $& $6.5 $& $2.5$\\  
\cline{2-8}
\multicolumn{8}{|c|}{} \\[-11.5pt]
\cline{2-8}
 & $20 $& & $16.5$& $15.1$& $5 $& $2.9 $&  $0.8 $\\  
\cline{2-2}\cline{4-8}
 & $50 $ & $1800 $&$13 $& $15.4 $& $5.3 $& $3.3 $& $1.2$\\  
\cline{2-2}\cline{4-8}
 & $100$ & (FDD) & $9.5$& $20 $& $7.8 $& $5.1 $& $1.9$\\  
\cline{2-8}
\multicolumn{8}{|c|}{} \\[-11.5pt]
\cline{2-8}
 & $20 $& & $15$& $2.4 $& $1.1 $& $0.57 $&  $0.15 $\\  
\cline{2-2}\cline{4-8}
 & $50 $& $3500 $& $13$& $2.8 $& $1.25 $& $0.68 $&  $0.24$ \\  
\cline{2-2}\cline{4-8}
 & $100 $& (TDD) & $10$& $3.4 $& $1.4 $& $1 $& $0.45 $\\  
\hline 
\end{tabular}
\label{tab:rate_UL}
\end{table*}

\
\section{Numerical Results}
\label{sec:NumericalResults}

\subsection{System Parameters}
Table~\ref{tab:SystemParmeters} lists the system parameters. In our work, a low, a medium, and a high number of active users are considered, i.e., $K\in \{20, 50, 100\}$. The users are distributed uniformly around the BS. Their heights are assumed at~$8$~m. Orthogonal frequency division multiplexing (OFDM) is employed with a cyclic prefix overhead of $5\%$. Considered carrier frequencies are $700$, $1800$, and $3500$~MHz with bandwidths of $10$~MHz, $20$~MHz, and $100$~MHz, respectively. FDD mode is used for $700$ and $1800$~MHz and TDD mode is used for $3500$~MHz as in the 5G new radio standard (NR)\cite{3GPPNR,falou2022enhancement}. 
 

\subsection{UL Rates Results}
In \cite{falou2022enhancement}, we computed the coverage range of the HTBS and the legacy BS, defined as the range for which more than $95\%$ of users have a DL data rate $\geq 10$~Mpbs. $d_\text{cov}$ represents the coverage distance, and users in a distance between $[0, d_\text{cov}]$ are covered. The DL transmit power is equal to $50$ dBm for HTBS and $46$ dBm for a legacy BS. As shown in Table~\ref{tab:rate_UL}, the ratio
\begin{equation}
\label{eq:ratio}
\frac{d_\text{cov}^\text{HTBS}}{d_\text{cov}^\text{Legacy BS}} \geq 5 ,
\end{equation}
with the same mMIMO configuration and for the same number of active users. Therefore, one HTBS covers an area more than $25$~times bigger than a legacy BS. More interestingly, when the number of active users is low, e.g., $K=20$, and for a low-frequency band, e.g., $f_c = 700$~MHz, the ratio in (\ref{eq:ratio}) can go to $7.5$ times, and the ratio between covered areas becomes $57$ times.  

To complete our study, we focus this paper on the UL. We compute the achievable UL data rate per user $\text{R}_{k}$, by more than 95\% of users, for several $\text{EIRP}_k$ values. The users are uniformly distributed in a circle centered at the BS with a radius equal to $d_\text{cov}$. The power ratio $\delta = 20$ dB and $\text{EIRP}_k$ represents the maximum allowed EIRP per user. The results are summarized in Table~\ref{tab:rate_UL}.

They show that a legacy BS is providing higher $\text{R}_{k}$ with respect to HTBS. This is directly related to their lower $d_\text{cov}$ and therefore their lower path loss. Furthermore, $\text{R}_{k}$ increases with the total number of active users $K$ as
\begin{enumerate}
    \item The coverage distance $d_\text{cov}$ decreases with the increase of $K$, decreasing the path loss,
    \item The ratio $M/K$ remains high, permitting the cancellation of most of the intra-cell interference. 
\end{enumerate}

In our HTBS cell, users inside the coverage range experience a DL rate  higher than or equal to $10$~Mpbs. The satisfaction of the users in the UL depends on their $\text{EIRP}_k$ and the system operating frequency. 
\begin{itemize}
    \item For $f_c = 700$ MHz, the rates $\text{R}_{k}$ are acceptable for $\text{EIRP}_k= 40$ and $33$ dBm and coverage area is maintained in both DL and UL. For $\text{EIRP}_k= 30$ and $23$ dBm, $\text{R}_{k}$ can be acceptable based on the target application and the traffic requirements. 
    \item For $f_c = 1800$ MHz, the similar conclusions to the case of $f_c = 700$ MHz can be drawn. The increase in path loss due to the increase in frequency is compensated by the increase in bandwidth from $W=10$ MHz for $f_c = 700$ MHz to $W=20$ MHz for $f_c = 1800$ MHz.
    \item For $f_c = 3500$ MHz, the UL rates $\text{R}_{k}$ are relatively low for all values of $\text{EIRP}_k$. This is due to the high path loss experienced by signals in higher frequencies. This increase in path loss is not compensated by an increase in the bandwidth, since the system is operating in a TDD mode and the DL part is predominant. 
\end{itemize}

High $\text{EIRP}_k$ values can be obtained either by the increase of the user transmit power and/or by the use of high-gain antennas. In the HTBS context, the use of a high-gain antenna is plausible as it is to be placed outdoors pointing toward the HTBS. Assuming a good backhaul link for the HTBS, users using high-gain antennas will experience an increase in their DL rates as well.  

\section{Techno-Economical Aspects for HTBS}
\label{sec:techno}
\subsection{Recommended Frequency Band}
Based on the results in Section~\ref{sec:NumericalResults} and in \cite{falou2022enhancement}, it is clear that the low-frequency band around $f_c = 700$ MHz is more suitable for HTBS deployment in rural areas as it provides higher coverage distances and good rates in both DL and UL. 

Rural areas are typically characterized by a low number of active users where the coverage is their main limitation. Many frequencies in the 5G NR lower bands can be considered, e.g., N12, N13, N14, N28, N29, and N71 bands \cite{3GPPNR}. The selected bandwidth $W$ can be increased to be higher than $10$ MHz depending on the requirements of the served location. 

In the sequel, we will quantify the number of covered users by an HTBS in the $700$ MHz frequency band, and their respective population densities. 

\subsection{Number of Covered Users and Population Density}
\begin{table*}
\normalsize \centering 
\caption{Number of Covered Users $N_\text{cov}$ and Population Density $\rho_\text{cov}$ by HTBS.}
\begin{tabular}{|c|c|c|c|c|c|} 
\hline
\rule{0pt}{10pt}    
$K$ &$d_\text{cov}$ [km] & $N_\text{cov}$ & $\rho_\text{cov}$ [$\text{pers.}/\text{km}^2$] & $\text{EIRP}_k$ [dBm] & $A_\text{cov}$ [$\text{km}^2$]  \\
\hline
\multicolumn{6}{|c|}{} \\[-13pt]
\hline
 $20$ & $37$ & $5,405$ & $1.26$ & $30$ &  $4,300$\\ 
\hline
 $50$ & $21$ & $13,513$ & $9.75$ & $30$ & $1,385$
 \\  
\hline
$100$ & $12.5$ & $27,027$ & $55$ & $23$ & $490$ \\  
\hline 
\end{tabular}
\label{tab:UL_cov}
\end{table*}
In our simulations, we considered three values for the number of simultaneously active users $K \in \{20, 50, 100\}$. This number represents the users that are sending data simultaneously at a one-time slot. In order to quantify the number of covered users, we proceed similarly to ~\cite{osoro2021techno,feltrin2021potential}. This number is important for network operators as it helps them to decide if HTBS deployment is lucrative on potential revenues. 

In~\cite{feltrin2021potential}, the authors considered the temporal aspects of the traffic for the case of rural Sweden. The traffic volume per user in the UL is of $1$~GB/month and $5$~GB/month in the DL.  The average traffic per user is, therefore, equal to $7.41$~kbps in the UL and $37.0$~kbps in the DL by assuming busy hours of $10$~hours/day. These values allow the approximation of the number of covered users per HTBS denoted by $N_\text{cov}$. 

Three coverage distances are obtained for HTBS  $d_\text{cov}^{K=20}=37$ km, $d_\text{cov}^{K=50}=21$ km, and $d_\text{cov}^{K=100}=12.5$ km as shown in Table~\ref{tab:rate_UL}.

For $d_{cov}^{K=100}=12.5$ km, the cell capacity is therefore $100\times10 \text{ Mbps}= 1$ Gbps. This corresponds to a number of served users of $N_\text{cov}^{K=100} = 1 \text{ Gbps} / 37.0 \text{ kbps} = 27,027$ users/cell. The percentage of active users from the total number of persons is $100/27,027 = 0.37\%$,  which is higher than typical values for rural areas, e.g.,  $0.1\%$ in \cite{osoro2021techno}. For this number of covered users, the required cell capacity in the UL is $27,027 \times 7.41  \text{ kbps}= 200$ Mbps. This capacity can be obtained with any $\text{EIRP}_k \geq 23$ dBm. For $\text{EIRP}_k = 23$ dBm the UL capacity is $100\times2.5 \text{ Mbps}= 250 \text{ Mbps} > 200$ Mbps. The average population density in this case is 
\begin{equation}
\label{eq:rho}
   \rho_\text{cov}^{K=100} = \frac{N_\text{cov}^{K=100}}{\pi {( d_\text{cov}^{K=100})}^2 } = \frac{27,027}{\pi 12.5^2} = 55 \text{ persons}/\text{km}^2.
\end{equation}

For one HTBS, Table~\ref{tab:UL_cov} provides the number of covered users $N_\text{cov}$ for the presented cases, their average population densities $\rho_\text{cov}$ in $\text{persons}/\text{km}^2$, the minimum required EIRP per user  $\text{EIRP}_k$, and the covered area $A_\text{cov}$.

\begin{figure}
    \centering
    \vspace{-1.2cm}
    \includegraphics[width=\columnwidth]{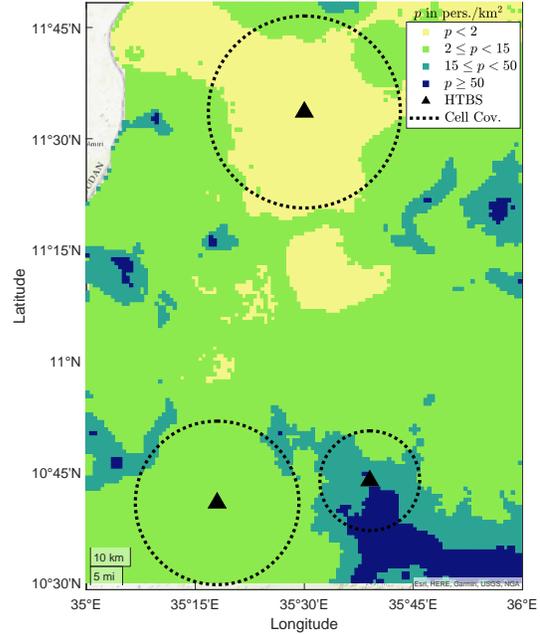}
    \vspace{-1.2cm}
    \caption{HTBS cell coverage in different rural locations.}
    \label{fig:EthUL}
\end{figure}

To further assess the practicability of the HTBS solution and the obtained $\rho_\text{cov}$, we consider a rural region in a country experiencing a high telecommunication imbalance such as Ethiopia \cite{zhang2021telecommunication}. This region is located between longitude $35^{\circ}$E-$36^{\circ}$E, and latitude: $10^{\circ}30'$N- $11^{\circ}45'$N with an area of $15,000 \text{ km}^2$. In Fig.~\ref{fig:EthUL}, we plot the population density $p$ in persons per $\text{km}^2$ for the considered region. These values are obtained from the Meta high-resolution population density maps \cite{Facebook}. We also provide three potential locations for the deployment of HTBSs corresponding to population densities close to $\rho_\text{cov}$ in Table~\ref{tab:UL_cov}. Fig.~\ref{fig:EthUL} clearly shows that the HTBS solution is suitable for rural regions characterized by low population densities as large areas can be covered by one HTBS. Furthermore, the cell coverage area decreases with the increase in population density.  

In practice, a duality exists between capacity and coverage. In villages and cities, HTBS can only be used in support of legacy BSs. For urban/suburban areas, connectivity is better provided with legacy BSs. As stated earlier, both types of BSs can co-exist guaranteeing a large coverage area when users are connected to the HTBS and high data rates when connected to a legacy BS.

\subsection{Available Low-Cost Locations for HTBS}
In our numerical results, we considered that the height of the HTBS is equal to $150$ m and the height of the users is $8$ m. We assumed that an effective height difference of $142$ m between the HTBS and the users is enough to guarantee that a LoS path exists. 

Available low-cost locations for HTBSs are:
\begin{enumerate}
\item \textit{TV Towers}: TV towers are powered and backhauled to broadcast served TV channels. One option is to benefit from TV towers by deploying an HTBS at the top of their towers' mast \cite{falou2022enhancement}. The TV service remains intact as the frequency used for data communication is not used by the TV service. 
\item \textit{High Hills}: We have shown that an HTBS can cover a large area. Contrary to legacy BSs that are usually deployed as close as possible to the users, an HTBS can be deployed at the highest location in the targeted area. The electrical powering can be done using renewable energy solutions such as solar panels and wind turbines, and the back-hauling can be done using a free-space optical (FSO) link backed-up with a microwave (MW) link.
\item \textit{Wind Turbines}: HTBS can also be deployed on available wind-turbines \cite{matracia2021exploiting}. Wind-turbine HTBSs are self-powered and back-hauling can also be done using FSO/MW links. In rural locations with good airflow, deploying wind-turbine HTBSs can solve both connectivity and power issues. 
\end{enumerate}

\section{Non-technological Challenges}
\label{sec:nontechno}
Several non-technological challenges for HTBS networks in rural regions can be listed~\cite{yaacoub2020key,feltrin2021potential,falou2022enhancement}. These challenges increase both CAPEX and OPEX, making their adoption decision not straightforward as:
\begin{enumerate}[label=(\roman*)]
    \item The lack of skilled engineers in rural locations is the major challenge for the proposed solution, especially in the least developed countries. These engineers are required not only for deployment but also for maintenance. 
     \item The poor infrastructure is also an important challenge. The deployment and maintenance of HTBSs require regular transportation of equipment. Bad roads make this task complicated.
    \item The absence or the regular interruption of electricity. Diesel generators or green power alternatives are required to power HTBSs.
\end{enumerate}

\section{Conclusion}
\label{sec:Conclusion}
In this paper, we have shown that HTBS with MU mMIMO can be considered a low-cost technique to provide connectivity in rural places. We provided the DL and UL data rates of an HTBS and showed that it can cover a large rural area, characterized by low population density, with respect to a legacy BS. Techno-economical aspects and non-technological challenges are also discussed.  Despite all these challenges, the HTBS solution remains one of the most cost-efficient solutions for rural connectivity with respect to other solutions such as HAPS, low earth orbit (LEO) satellites, and UAV-BSs. 


\bibliographystyle{IEEEtran}
\bibliography{IEEEabrv,Bibliography}


\end{document}